# Superconducting Antenna Concept for Gravitational Wave Radiation

A. Gulian[a*,1], J. Foreman[b], V. Nikoghosyan[a,c], L. Sica[a], J. Tollaksen[a,d], and S. Nussinov[a,e]

[a]*Chapman University, Institute for Quantum Studies, Orange CA, 92866 & Burtonsville MD, 20866, USA*
[b]*Independent Researcher, Alexandria VA, 22310, USA*
[c]*Physics Research Institute, National Academy of Sciences, Ashtarak, 0203, Armenia*
[d]*Chapman University, Schmid College for Science and Technology , Orange CA, 92866*
[e]*School of Physics and Astronomy, Tel-Aviv University, Ramat-Aviv, 69978, Tel-Aviv, Israel*

*Abstract* – We present an idea for detecting gravitational waves (GWs) by measuring the current induced in a superconducting loop by the rotation of a frame to which it is attached. The frame experiences a torque caused by a GW propagating perpendicular to it because it is made of materials of different density in adjacent quadrants. Superconducting Cooper pair system responds symmetrically to the wave and stays at rest, while the ions of the superconductor are being accelerated by the moving frame. This generates an electric current in the loop which can be measured precisely by sensitive electronics. For that task the antenna consists of two superconducting loops parallel to each other. One of them, the primary loop, rotationally oscillates because of the described reasons, while the second stays at rest.  In the oscillating loop the current creates an oscillating magnetic flux. This flux should be compensated due to fluxoid quantization, by the oppositely directed current in the non-moving, secondary loop, in which the ions are at rest, and the Cooper pairs move. We estimate the resulting signal/noise ratio and discuss the signal detectability. Several designs are envisioned, both for terrestrial and for orbital arrangements of the antenna.

1. **Introduction**

The struggle of the physics community to observe GWs is ongoing, and requires continuing intellectual efforts [1-6]. Since the energy exchange between the source and an antenna [7] is proportional to the mass of the antenna, as well as to the square of its lateral size, detector-antennas with large masses [8] and long baselines [9] are at present the preferred candidates [10]. However, signal/noise characteristics of detectors are critically important for successful observation of GWs, and from this perspective detectors utilizing Bose-Einstein condensates (BEC) or their relatives, Cooper pair condensates (CPC), may have an unsurpassed performance. Since photons do not constitute a BEC, optical interferometric devices cannot provide the benefit of practically infinite coherence time, while by contrast the current in a superconducting ring, though in a metastable state, can flow for a time longer than the age of the Universe. This is because thermodynamic noise cannot degrade the phase coherence of a BEC or CPC. Small disturbances in the density of the condensate will be "healed" by the condensate itself, thus making the coherence time as long as required. By contrast, a photonic ensemble may be degraded one photon at a time, which yields finite coherence time for photon ensembles, and which should be compensated by a laser inputting more photons and thus adding inevitable noise to the system. Speaking of noise, superconductors have single-particle excitations (electrons and holes) separated by an energy gap from the CPC, and subject to thermodynamic noise. Charge neutrality couples the motion of CPC and single-particle excitations. However, at sufficiently low temperatures, because of this energy gap, the number of single-particle excitations is exponentially small, and the perturbing effect on the CPC via this coupling becomes negligible.

Low noise is an invaluable asset when dealing with small signals. In this article we describe a concept which, if properly implemented, has chances of enabling the detection of GW radiation with orders of magnitude better sensitivity than the existing and developing detectors. We demonstrate this statement for periodic sources of GW with given frequency. Generalization for arbitrary sources is straightforward, but is beyond the scope of this paper.

## 2. How to initiate electric current in a superconductor by a GW field

Consider a GW incident on four free masses arranged in a quadrupolar fashion in free space. Figure1 illustrates the quadrupolar nature of the forces caused by a "+" polarized GW propagating perpendicular to the plane of the drawing. Since a rotation of the GW source by 45 degrees will transform "+"-polarization into an "x"-polarization, it is enough to consider detectors only for the "+" polarization of the GW. In case of arbitrary polarization, one can rotate the detector to the most sensitive position.

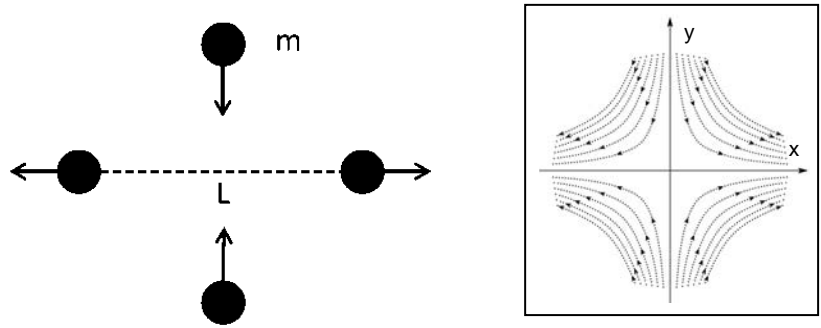

FIG.1. *Left panel*: Tidal action of "+" polarized GW on a quadrupolar arrangement of free particles. The GW is incident perpendicular to the plane. The corresponding GW force field lines [11] are indicated in the *right panel*.

Assuming this position, the plane GW with amplitude $h = h_0 \sin \omega t$ will adiabatically impart to and retrieve from the masses in Fig. 1 a kinetic energy $E_{kin} = m_{tot}(L\omega h_0)^2/2, \quad m_{tot} = 4m$. The question is: how can this kinetic energy, or at least a part thereof, be converted into a current? A set-up addressing this question is shown in Fig. 2.

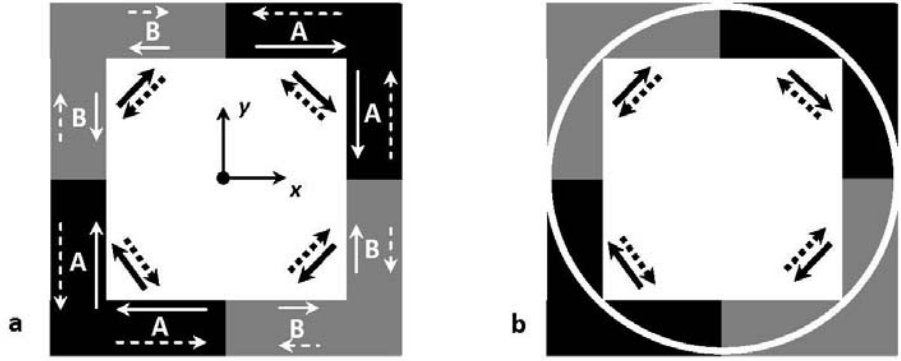

FIG.2. a) A frame made of materials A and B with different densities ($\rho_A > \rho_B$). The rotation of the frame around its center of mass (small black dot) always follows the direction prescribed by A. The solid white arrows indicate the driving force direction during the first half-period, and the dashed white arrows during the second half-period of the GW. The black arrows at the corners indicate the direction of the frame rotation.
b) A superconducting loop rigidly attached to the frame is indicated by the white circle.

During the first half-period, the *x*-components of the GW forces stretch the horizontal bars pushing the upper halves to the right and the lower halves to the left, and the forces along the *y*-axis contract the two vertical bars by pushing the right halves down and the left halves up. The same pattern, but with all forces and resulting motions reversed, occurs in the second half-period. If $\rho_A$ is the same as $\rho_B$, then all forces would identically cancel and there would be no torque. However, in the case where $\rho_A > \rho_B$ the rotation of the frame around its center of mass always follows the direction prescribed by A. This explains how to obtain a circular oscillation of a mechanical system from a curl-free GW field. One can recall a familiar analog in optics: filtering linearly polarized light by appropriate crystals can convert it into circular polarization – which carries net angular momentum along the direction of propagation– leaving atoms with the opposite angular momentum.

Next we wish to generate a current from this oscillatory circular motion. To this end, we rigidly attach a superconducting circular loop onto the frame, as shown in Fig. 2b. Then ions of superconducting material will move with the frame, while the Cooper pairs will stay at rest. As a result, a magnetic flux will be generated by the rotating loop. If a second similar superconducting loop is suspended in a plain parallel to the first, being mechanically detached from it, and secured to stay at rest, an opposite current will arise in this loop, so that the flux will be fully compensated (flux transformer based on fluxoid quantization). We will now evaluate the electronic current caused by the GW in these loops.

### 3. Estimates for the current

For simplicity, we consider the mass of the moving superconductor loop to be negligibly small compared to the mass of the frame. To simplify the picture even more, we will consider the case where $\rho_A >> \rho_B$. Then the linear velocity of rotation of the loop of radius *L* in response to the GW is $v \sim L(dh/dt) \sim L\omega h_0$ relative to the laboratory reference system. In this laboratory reference system the Cooper pairs stay at rest, and the ions of the superconductor move and constitute a current. Correspondingly, for an observer moving with the ions, the Cooper pairs move through the loop and constitute a current. Defining the effective cross section of current flow as $s_{eff}$, and assuming constant current density $j$, then oscillatory current $I = I_0 \cos\omega t$ has amplitude:

$$I_0 \sim j s_{eff} = en\mathrm{v}\, s_{eff} \sim enL\omega h_0 s_{eff}, \qquad (1)$$

where $n$ is the density of electric carriers in the superconducting material. It is advantageous to express this response in terms of the magnetic flux quantum $\phi_0$. A current $I$ in a loop of radius $R$ creates a magnetic field **B** which at a distance $\alpha$ from the loop axis in the plane of the loop has amplitude

$$B(\alpha) = \frac{\mu_0 I_0}{2\pi(R+\alpha)}\left\{K(k_0) + \frac{R+\alpha}{R-\alpha}E(k_0)\right\}, \quad k_0 = 2\frac{\sqrt{R\alpha}}{R+\alpha}, \qquad (2)$$

where $K(k)$ and $E(k)$ are complete elliptic (Legendre) functions of the first and second kind [12]. Integrating this, one finds the total flux in the loop:

$$\Phi(I_0) = 2\pi \int_0^R B(\alpha)\alpha\, d\alpha = \beta \cdot I_0 \qquad (3)$$

where $I_0$ is expressed in amperes, and for $R=L=10m$ one obtains a numeric value $\beta \approx 5\cdot 10^{10}\cdot\phi_0/Amp$. This flux could be measured by coupling it via a flux transformer to a SQUID pick-up loop using standard methods of superconducting electronics (e.g., [13,14]). For that task a second loop should be placed in parallel to the first one, which can be regarded as the secondary loop of the transformer. This loop can be at rest in the laboratory system, which simplifies its coupling to a SQUID and other electronics. Commonly available SQUIDs routinely achieve a noise floor $\delta\Phi \sim 10^{-6}\,\phi_0/Hz^{1/2}$ at 4.2K [15]. Even better results can be expected at much lower temperatures [16] which could be used if required. Assuming this 4K operational temperature and adopting $\delta\Phi \sim 10^{-6}\,\phi_0/Hz^{1/2}$ for determining the detection threshold (defined as signal/noise=1), we obtain from Eqs. (1)-(3) a strain sensitivity $h_0/(\delta f^{1/2}) \sim 10^{-26}\ Hz^{-1/2}$ at $\omega \sim 360\sec^{-1}$ [17], using $n \sim 10^{30} m^{-3}$ and $s_{eff} \sim 10^{-5} m^2$.

Because of the Meissner effect, low frequency currents in superconducting wires flow only within a surface layer with thickness equal to the London penetration depth $\lambda_L$. Thus, for example, a *3mm* wire is not sufficient to achieve the $s_{eff}$ used above. To obtain a cross section as large as the specified $s_{eff}$, one may surround the whole toroidal frame, which could have a diameter $d \geq 30cm$, by a corrugated surface as shown in Fig. 3.

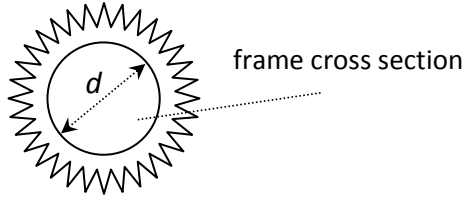

FIG. 3. Cross sectional view of corrugated surface of the wire surrounding the frame (not in scale) for enhancing $s_{eff}$.

Then for $d \approx 0.3 m$, $\lambda_L \approx 10^{-7} m$, and corrugation enhancement factor $\eta \approx 3 \cdot 10^2$, we arrive at the value of $s_{eff} \approx \eta d \lambda_L$ used above; these parameters still allow some flexibility, and if required may be further adjusted for even a better outcome.

### 4. Inherent thermodynamic noise of the antenna

In addition to the above-mentioned noise of signal-registering electronics, other noise factors should be taken into account. In particular, intrinsic thermodynamic noise (of Johnson-Nyquist origin) generated by electron-hole excitations (mentioned above) is inevitable in superconductors. This noise contribution can be modeled via a finite resistor attached in parallel to the superconducting wire. Then:

$$\langle I_{noise} \rangle = [4(k_B T / R_n) \delta v]^{1/2}, \qquad (4)$$

where $k_B$ is the Boltzmann constant, and $R_n$ is the resistance of the normal component of the superconductor:

$$R_n = (\rho L / S) \exp[\Delta /(k_B T)]. \qquad (5)$$

In Eq. (5) $S$ is the wire cross section, $\rho$ is the resistivity of unpaired electrons in a superconductor, and $\Delta = \Delta(T)$ is the BCS gap in their spectrum [13,18]. This current (4) creates a fluctuational flux in the loop in accordance to Eq. (3):

$$\langle \delta \Phi_{noise} \rangle \approx 5 \cdot 10^{10} \cdot \phi_0 \cdot \{\exp[-\Delta /(2k_B T)] \cdot [4(k_B T S / \rho L) \delta v]^{1/2} (1/ Amp)\}. \qquad (6)$$

We would like this noise not to exceed the noise introduced by the SQUID detector, that is $\langle \delta \Phi_{noise} \rangle \leq 10^{-6} \cdot \phi_0 (\delta v / Hz)^{1/2}$. This requirement is possible to satisfy if the following condition occurs:

$$\exp[-\Delta /(2k_B T)] \cdot [4(k_B T S / \rho L)]^{1/2} < 2 \cdot 10^{-17} (Amp / Hz^{1/2}) \qquad (7)$$

The condition (7) is easier to satisfy if $S = s_{eff}$. Using the values for $s_{eff}$ and $L$ specified above, and $\rho \sim 10 \mu\Omega \cdot cm$, $\Delta \sim 10 meV$ and operational temperature $T \sim 4K$, we have for the l.h.s. of inequality (7) an estimate: $\exp[-\Delta/(2k_B T)] \cdot [4(k_B T s_{eff} / \rho L)]^{1/2} \approx 10^{-17}$ $(Amp/Hz^{1/2})$, which means that condition (7) is satisfied by this choice of parameters. Moreover, superconductors with even larger values of normal resistivity and larger energy gaps are readily available. This generally means that the estimate for the antenna strain sensitivity

$$h_0 \sim 10^{-26} \, (\delta f / Hz)^{-1/2} \tag{8}$$

derived in Section 3 by taking only the SQUID noise into account, stays valid when the inherent thermodynamic noise of the antenna is also taken into account.

## 5. Discussion and conclusions

In practice, our free-frame motion based device appears to be most appropriate for satellite-based implementation. However, as shown in Fig. 4, a multi-toroidal design may be realized for working in terrestrial conditions as well.

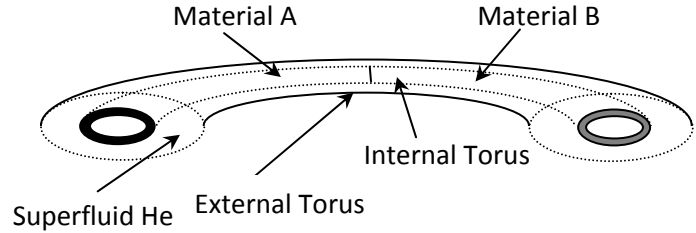

FIG.4. Toroidal design of GW antenna for terrestrial operation: superfluid helium fills the space between the external and internal tori. The internal torus is made hollow for buoyancy. It consists of materials A and B (only two are shown). Primary loop of the superconducting wire is located on the surface of the internal torus, while the detecting loop is on the external one (both are not shown).

In this design the (corrugated) surface of the internal torus (the frame) is homogeneously covered by a thin superconducting layer, so that as above, the Cooper pairs are moving relative to the ionic lattice when the internal torus is actuated by the GW. The external torus will carry a magnetic flux pick-up loop. Noise factors inherent to this arrangement should be considered separately.

To summarize, we suggested a way of converting the transverse acceleration of masses in the field of periodic quadrupolar GWs into an oscillatory rotational motion of oppositely charged particles relative to each other. Relative motion of oppositely charged particles generates detectable magnetic flux in superconducting circuits, and yields a conceptually new GW antenna with potentially phenomenal sensitivity. In this article we had neither a goal to provide a detail engineering design for this antenna, nor an intension to compete with known mature technologies. A considerable engineering effort will be required to produce a working model from the ideas presented here, and other noise sources remain to be considered. However, given the extreme importance and difficulty of the general task of GW detection, we believe that our approach should be further analyzed in sufficient depth to enable practical implementations.

*Acknowledgements.* One of us (AG) would like to express his gratitude to L.P. Grishschuk, P.B. Abramian-Barco, and A.E. Sargsyan for enlightening and encouraging discussions at the early stage of this work.